\documentclass{article}

\newcommand{\bb}{\begin{eqnarray}}
\newcommand{\ee}{\end{eqnarray}}
\begin{document}
\title{Chiral phase in fermionic measure and the resolution of the strong CP problem}
\author{P. Mitra\thanks{mitra@tnp.saha.ernet.in}\\
Saha Institute of Nuclear Physics,\\ 1/AF Bidhannagar,\\
Calcutta 700064, India}
\date{hep-th/0104186}
\maketitle

\begin{abstract}{The fermionic measure in the functional integral 
of a gauge theory suffers from an ambiguity
in the form of a chiral phase. By fixing it, one is led
once again to the conclusion that a chiral phase in the quark mass term of QCD
has no effect and cannot cause CP violation.}
\end{abstract}

\section*{}
At present, a popular approach to anomalies is in the functional integral formalism. 
Here an integration is carried out over different field configurations.
The integrand involves the classical action which possesses all classical symmetries,
but the measure of integration may break some of these \cite{fujikawa},
and lead to anomalies, which can be calculated if a regularization is employed.
The measure of integration over fermions has been extensively studied in connection with
chiral anomalies, but as will be demonstrated here, there is still a certain ambiguity
left. This ambiguity relates to the case when the fermion mass is non-vanishing,
and is practically relevant in the important context of QCD. 
In the standard model, the parity-violating electroweak sector 
gives rise to an unknown chiral
phase in the quark mass term of QCD, and this has been at the basis of 
the strong CP problem. The ambiguity in the fermion
measure would imply an ambiguity in the amount of CP violation caused by the chiral
phase in the fermion mass term. The natural way of fixing the ambiguity makes the 
CP violation
disappear. This is in consonance with the results obtained by 
explicit regularization of
the action, namely the generalized Pauli-Villars regularization \cite{bcm} 
and lattice regularizations \cite{lat}.
This reconfirms the basic idea behind the solution of the strong CP problem
in the more formal framework in which anomalies arise primarily from the measure and the 
regularization is relegated to the background.  
Indeed, this investigation was motivated by the fact that the earlier discussions
resolving the strong CP problem were restricted to regularized
actions and trivial measures, whereas many of the misconceptions that made up the strong
CP problem arose in analyses using nontrivial measures instead of regularizations.

For a theory like quantum chromodynamics, with fermions interacting vectorially
with gauge fields, the euclidean action may be written as
\bb
S=\int \bar\psi(\gamma^\mu D_\mu-m)\psi +S_g,
\ee
where $D$ is the covariant derivative, hermitian $\gamma$-matrices are used
and $S_g$ stands for the gauge field action. 
The fermionic integration is supposed to be over $\psi, \bar\psi$, 
but in the context of anomalies, these are conventionally expanded in 
orthonormal eigenfunctions of $\gamma^\mu D_\mu$ \cite{fujikawa},
\bb
\psi=\sum_n a_n\phi_n,\quad \bar\psi=\sum_n \bar{a}_n\phi^\dagger_n,
\label{zero}\ee
and the expansion coefficients $a_n, \bar{a}_n$, which are Grassmann variables, 
are integrated over. The functional integral is
\bb
Z=\int {\cal D}A \prod_n \int da_n \prod_n \int d\bar{a}_n e^{-S}.
\label{Z}\ee
A chiral transformation of the fermion fields
\bb
\psi\rightarrow e^{i\alpha\gamma^5}\psi,\quad
\bar\psi\rightarrow\bar\psi e^{i\alpha\gamma^5},
\label{chi}\ee
causes $a_n,\bar{a}_n$ to change, 
\bb
a_n&\rightarrow&\sum_m\int\phi_n^\dagger e^{i\alpha\gamma^5}\phi_m a_m,\nonumber\\
\bar{a}_n&\rightarrow&\sum_m\bar{a}_m\int\phi_m^\dagger e^{i\alpha\gamma^5}\phi_n,
\label{J}\ee
and there is a nontrivial Jacobian 
which can be calculated with an appropriate regularization. It is this Jacobian 
which is responsible for the axial anomaly in this approach \cite{fujikawa}
because the action is 
chirally invariant when $m=0$. When $m\neq 0$, the divergence of the axial current
gets a contribution from the explicit breaking due to this mass in addition to
the anomaly.

We are interested here in the situation where the fermion mass term has a chiral phase.
Let us write the action as 
\bb
S_{\theta'}=\int \bar\psi(\gamma^\mu D_\mu-me^{i\theta'\gamma^5})\psi +S_g,
\ee
where the prime in the symbol $\theta'$ is used to distinguish it from the coefficient
of the $F\tilde F$ term in QCD. The corresponding functional integral will be given by an
equation like (\ref{Z}), but what are the corresponding $a_n,\bar{a}_n$? One may
na{\"\i}vely follow \cite{fujikawa} and use (\ref{zero}). In the presence of the
chiral phase in the mass term, we shall distinguish
these expansion coefficients by the superscript 0, writing 
\bb
\psi=\sum_n a^0_n\phi_n,\quad \bar\psi=\sum_n \bar{a}^0_n\phi^\dagger_n,
\ee
and the functional integral is
\bb
Z_0=\int {\cal D}A \prod_n \int da^0_n \prod_n \int d\bar{a}^0_n e^{-S_{\theta'}}.
\ee

One may, instead, try to {\it derive} a functional integral 
in the presence of a chiral phase
in the fermion mass term from the simple case where there is no such phase.
$S_{\theta'}$ is obtained from $S$ by a chiral transformation (\ref{chi}) with
$\alpha=\theta'/2$, so the expansion (\ref{zero}) gets altered to
\bb
\psi=e^{-i\theta'\gamma^5/2}\sum_n a^{\theta'}_n\phi_n,\quad \bar\psi=\sum_n 
\bar{a}^{\theta'}_n\phi^\dagger_n e^{-i\theta'\gamma^5/2}.
\label{theta'}\ee
The chiral phase sticking out on the right hand side may look
unfamiliar and strange, but it has to be remembered that the basis
of expansion is not {\it a priori} fixed \cite{jog}, and the phase can be thought of as
causing a change of basis. Correspondingly
\bb
Z_{\theta'}=\int {\cal D}A \prod_n \int da^{\theta'}_n \prod_n 
\int d\bar{a}^{\theta'}_n e^{-S_{\theta'}}.
\ee
This integral is different from the previous one because of the anomaly: A
redefinition of the expansion coefficients along the lines of (\ref{J}), followed 
by a calculation of the Jacobian can be used to convert 
$a^{\theta'}_n, \bar{a}^{\theta'}_n$ to
$a^0_n,\bar{a}^0_n$, but an $F\tilde F$  term gets added to the action, showing the 
difference between $Z_0, Z_{\theta'}$. $Z_0$ violates CP, while $Z_{\theta'}$ does not,
because it is equivalent to the case with no chiral phase in either action or measure.

This already shows the ambiguity in the fermion measure, but one can be even more
general. Note that $Z_0$ has been defined by postulating the standard measure in
the presence of $\theta'$, and $Z_{\theta'}$ has been derived by falling back on the
standard measure only in the absence of any chiral phase in the mass term. 
There is a continuum of
possibilities in between: one can start with a mass term having some chiral phase 
different from zero and $\theta'$, and postulate the standard measure in that situation.
Then one makes a chiral transformation to reach $\theta'$ in the mass term, but
this chiral transformation alters the expansion of the fields. Thus, in general, 
\bb
\psi=e^{-i\beta\gamma^5/2}\sum_n a^\beta_n\phi_n,\quad \bar\psi=\sum_n 
\bar{a}^\beta_n\phi^\dagger_n e^{-i\beta\gamma^5/2},
\label{beta}\ee
and the functional integral is
\bb
Z_\beta=\int {\cal D}A \prod_n \int da^\beta_n \prod_n \int d\bar{a}^\beta_n 
e^{-S_{\theta'}}.
\ee
Here $\beta$ is a real parameter.
$Z_0,Z_{\theta'}$ correspond to the special cases $\beta=0,\theta'$ respectively.
This functional integral depends on $\beta$ because of the nontrivial
$\beta$-dependent Jacobian which would be
involved in removing the chiral phase from the expansion 
by redefining the expansion coefficients: one would get
a $\beta$-dependent $F\tilde F$ term added to the action.  
This $\beta$-dependence in the expansion, which can be converted to a
$\beta$-dependence in the action, is the ambiguity referred to above.

This new parameter $\beta$ may appear unnecessary because it has been
overlooked earlier, but it can be nonzero, and if it is, it changes all old
discussions of strong CP physics. The standard picture used to be that there
are two parameters: one in the chiral phase of the fermion mass term, and the
other in the $F\tilde F$ term hidden here in $S_g$. By chiral transformations
one would then try to show that physical quantities depend on the algebraic
combination of these two phases. What we are pointing out here is that the
measure accommodates a third phase, and it is the algebraic combination of
these {\it three} phases that is relevant for physics.  A third phase also
arises if explicit regularizations are used as in \cite{bcm,lat} and has the
same significance in the context of strong CP physics. It may be recalled that
that phase gets fixed by the desire to maintain the classically conserved
parity (see below) instead of running into regularization artefacts.

One may ask whether the anomaly depends on the parameter $\beta$. It is easy 
to see that
even in the presence of a chiral phase in the expansion as in (\ref{beta}), a chiral
transformation (\ref{chi}) leads to the change (\ref{J}) of the expansion
coefficients, and the anomaly is standard.
 
Can one then choose between the different values of $\beta$? 
A choice is already suggested
by the way the alternatives were introduced above. $Z_0$
and the more general $Z_\beta$ for $\beta\neq\theta'$ were obtained by postulating the
standard measure in an {\it unknown} domain, involving the presence of a chiral phase, 
while $Z_{\theta'}$, corresponding
to $\beta=\theta'$ was derived from the standard measure in the {\it known}
domain where there is no chiral phase, which can probably be trusted more. 
In view of the conflict, this may be the functional integral to be
adopted.

Another way of choosing the parameter $\beta$ is through the consideration of
parity, as in \cite{bcm,lat}. As explained there, the classical theory with a
chiral phase in the mass term does {\it not} violate parity, in spite of
appearances to the contrary. One simply has to redefine the parity operation
for fermions. As it is, the parity operation for fermions in the absence of
a chiral phase in the mass term involves a $\gamma^0$. In the presence of the
chiral phase $\theta'$, this is altered and the appropriate parity operation is
\bb
\bar\psi(x_0,\vec x)&\rightarrow &\bar\psi(x_0,-\vec x)
e^{i\theta'\gamma^5}\gamma^0\nonumber\\
\psi(x_0,\vec x)&\rightarrow &
\gamma^0e^{i\theta'\gamma^5}\psi(x_0,-\vec x).
\ee
This change is justified by the fact that the changed transformation,
together with the standard parity operation for gauge fields, leave $S_{\theta'}$
invariant, except to the extent that there is an 
$F\tilde F$ term in $S_g$, which explicitly violates parity. 
As $S_{\theta'}$ has this symmetry, it is natural to try to see if the
fermion measure can be arranged to have the same symmetry. If not, it would be a case
of an anomaly. Fortunately, there is no parity anomaly: the measure preserves the
symmetry, {\it i.e.} the above transformation leaves $a^\beta_n,\bar{a}^\beta_n$ in
(\ref{beta}) invariant if and only if $\beta=\theta'$. 
This can be seen by noting that under the parity operation,
\bb
\phi_n(x_0,\vec x)\rightarrow\gamma^0\phi_n(x_0,-\vec x),
\ee
so that
\bb
\phi_n^\dagger(x_0,\vec x)e^{-i\beta\gamma^5/2}&\rightarrow&
\phi_n^\dagger(x_0,-\vec x)e^{-i\beta\gamma^5/2}
e^{i\beta\gamma^5}\gamma^0,\nonumber\\
e^{-i\beta\gamma^5/2}\phi_n(x_0,\vec x)&\rightarrow&\gamma^0
e^{i\beta\gamma^5}e^{-i\beta\gamma^5/2}\phi_n(x_0,-\vec x).
\ee
This  confirms that one has to take $\beta=\theta'$.
If one uses $\beta\neq\theta'$, and in particular the value zero,
one is guilty of violating parity by a deliberate choice of measure when a
parity conserving measure is available. Parity violation is then an
artefact of the measure.
 
Unfortunately, the chiral phase $\theta'$ in the expansion has been overlooked in the
literature and this has caused immense confusion. In particular, if a chiral
transformation is carried out to remove $\theta'$ 
from the action, it automatically goes away from the expansion relations connecting the
fermion fields with the eigenfunctions of the Dirac operator. No Jacobian
needs to be calculated, and no $F\tilde F$ term gets generated in the 
action \cite{baluni}.   
This means that the chiral phase has no effect,
in agreement with the recent resolution of the strong CP problem
\cite{bcm,lat} through similar demonstrations that a chiral phase in the quark mass
term does not violate CP. It is of interest to note that in all cases a chiral
phase had to be identified to absorb the chiral phase in the mass term. In the
other cases, this absorbing chiral phase appeared in the regulators, while in
the present one it appears in the expansion or equivalently in the measure. 

\bigskip

The problem studied here arose out of
discussions with Haridas Banerjee on the strong CP problem.

\end{document}